\documentclass[twocolumn,showpacs]{revtex4}
\usepackage{graphicx}
\usepackage{dcolumn}
\usepackage{bm}
\usepackage{amsmath}
\usepackage{amsfonts}
\usepackage{amssymb}

\providecommand{\U}[1]{\protect\rule{.1in}{.1in}}

\begin{document}

\title{Resonant quantum kicked rotor with two internal levels}
\author{Guzm\'an Hern\'andez and Alejandro Romanelli}
\altaffiliation{alejo@fing.edu.uy}
\affiliation{Instituto de F\'{\i}sica, Facultad de Ingenier\'{\i}a\\
Universidad de la Rep\'ublica\\
C.C. 30, C.P. 11300, Montevideo, Uruguay}
\date{\today }

\begin{abstract}
We develop a system consisting of a quantum kicked rotor with an additional
degree of freedom. This models a single two-level atom with internal ground
and excited states, and it is characterized by its quantum resonances with
ballistic spreading and by the entanglement between the internal and
momentum degrees of freedom. These behaviors establish an equivalence
between our model and the usual quantum walk on the line.
\end{abstract}

\maketitle

\section{Introduction}

Advances in technology during the last decades have made it possible to
obtain samples of atoms at temperatures in the $nK$ range \cite{Cohen}
(optical molasses) using resonant or quasiresonant exchanges of momentum and
energy between atoms and laser light. The experimental progress that has
allowed to construct and preserve quantum states has also opened the
possibility of building quantum computing devices \cite%
{Dur,Sanders,Du,Berman} and has led the scientific community to think that
quantum computers could be a reality in the near future. This progress has
been accompanied with the development of the interdisciplinary fields of
quantum computation and quantum information. In this scientific framework,
the study of simple quantum systems such as the quantum kicked rotor (QKR)
\cite{Casati0,Izrailev} and the quantum walk (QW) \cite{Kempe} may be useful
to understand the quantum behavior of atoms in optical molasses.

The QKR is considered as the paradigm of periodically driven systems in the
study of chaos at the quantum level \cite{Casati0}. This system shows
behaviors without classical equivalent, such as quantum resonance and
dynamical localization, which have posed interesting challenges both in
theoretical and experimental \cite{Nielssen} terms. The occurrence of
quantum resonance or dynamical localization depends on whether the period of
the kick $T$ is a rational or irrational multiple of $4\pi$. For rational
multiples, the behavior of the system is resonant while for irrational
multiples the average energy of the system grows in a diffusive manner for a
short time and then the diffusion stops and localization appears. From a
theoretical point of view the two types of values of $T$ determine the
spectral properties of the Hamiltonian. For irrational multiples the energy
spectrum is purely discrete and for rational multiples it contains a
continuous part. Both resonance and localization can be seen as interference
phenomena, the first being a constructive interference effect and the second
a destructive one. The QKR has been used as a theoretical model for several
experimental situations dealing with atomic traps \cite%
{Moore0,Kanem,Chaudhury,Moore1,Robinson0,Robinson1,Bharucha,Oskay} and is a
matter of permanent attention \cite%
{Schomerus0,Schomerus1,alejo,alejo0,alejo1,alejo2,alejo3,alejo4,alejo5}.

The quantum walk has been introduced \cite%
{Aharonov,Meyer,Watrous,Ambainis,Kempe,Kendon1,Kendon2,Konno,Salvador} as a
natural generalization of the classical random walk in relation with quantum
computation and quantum information processing. In both cases there is a
walker and a coin; at every time step the coin is tossed and the walker
moves depending on the toss output. In the classical random walk the walker
moves to the right or to the left, while in the QW coherent superpositions
right/left and head/tail happen. This feature endows the QW with outstanding
properties, such as the linear growth with time of the standard deviation of
the position of an initially localized walker. as compared with its
classical counterpart, where this growth goes as $t^{1/2}$. This has strong
implications in terms of the realization of algorithms based on QWs and is
one of the reasons why they have received so much attention. It has been
suggested \cite{Childs} that the QW can be used for universal quantum
computation. Some possible experimental implementations of the QW have been
proposed by a number of authors \cite%
{Dur,Travaglione,Sanders,Knight,Bouwmeester,Do,Chandrashekar}. In particular
the development of techniques to trap samples of atoms using resonant
exchanges of momentum and energy between atoms and laser light may also
provide a realistic frame to implement quantum computers \cite{Cirac}.

A parallelism between the behavior of the QKR and a generalized form of the
QW was developed in Refs. \cite{alejo0,alejo1} showing that these models
have similar dynamics. In those papers, the modified QW was mapped into a
one-dimensional Anderson model \cite{Anderson}, as had been previously done
for the QKR \cite{Grempel}. In the present paper, following the work of
Saunders $et$ $al.$ \cite{Saunders1,Saunders2} we propose a modification of
the QKR. We study some properties of this new version of the QKR and
establish a novel equivalence between this new QKR and the QW. Essentially,
the new QKR has an additional degree of freedom which describes the internal
ground and excited states of a two-level atom. %This new degree of freedom
%establishes a further topological equivalence between the QKR and the QW.
We call this new system the two-level quantum kicked rotor (2L-QKR). In this
system the internal atomic levels are coupled with the momentum of the
particle. This coupling produces an entanglement between the internal
degrees of freedom and the momentum of the system.

The rest of the paper is organized as follows, in the next section we
present the 2L-QKR system. In the third section we obtain the time evolution
of the moments. In the fourth section the entanglement between the internal
degrees of freedom and momentum is studied. In the last section some
conclusions are drawn.

\section{Two-level quantum kicked rotor}

We consider a Hamiltonian that describes a single two-level atom of mass $M$
with center-of-mass momentum described by the operator $\widehat{P}$. Its
internal ground state is denoted by the vector $|g\rangle$ and its excited
state by the vector $|e\rangle$. The internal atomic levels are coupled by
two equal-frequency laser traveling waves with a controllable phase
difference. Following \cite{Saunders1}, after a shift of the energy values,
the 2L-QKR Hamiltonian can be written as
\begin{eqnarray}
\widehat{H} &=&\frac{\widehat{P}^{2}}{2M}+\hbar \Delta |e\rangle \langle e|
\notag \\
&&+K\delta _{T}(t)\cos (k_{L}\widehat{z})(|e\rangle \langle g|+|g\rangle
\langle e|).  \label{ec_ham_qkr_2niveles}
\end{eqnarray}
Here $\Delta $ is the detuning between the laser frequency and atomic
transition frequency. $K$ is proportional to the Rabbi frequency and we
shall refer to it as the strength parameter.
\begin{equation}
\delta _{T}(t)=\sum_{n=0}^{n=\infty }\delta (t-nT)  \label{Dirac}
\end{equation}%
is a series of periodic Dirac's delta applied at times $t=nT$ with $n$
integer and $T$ the kick period. $\widehat{z}$ is the operator of the atom's
center of mass position. Finally, $k_{L}$ is the laser wave-vector magnitude
along the $z$ direction.

Unlike the QKR, in the 2L-QKR the conjugate position and momentum operators
have discrete and continuous components, \emph{i.e.}
\begin{equation}
\widehat{z}=\frac{1}{k_{L}}(2\pi \widehat{l}+\widehat{\theta })  \label{z}
\end{equation}%
\begin{equation}
\widehat{P}=\hbar {k_{L}}(\widehat{k}+\widehat{\beta })  \label{p}
\end{equation}%
where the eigenvalues of $\widehat{l}$ and $\widehat{k}$ are integers and
the eigenvalues of $\widehat{\theta }\in \lbrack -\pi ,\pi )$ and the
eigenvalues of the quasimomentum $\widehat{\beta }\in \lbrack -1/2,1/2)$. It
is important to point out that the operator $\widehat{\beta }$ commutes with
both $\widehat{k}$ and $\widehat{\theta }$. Using Eqs.(\ref{z},\ref{p}) to
substitute $\widehat{z}$ and $\widehat{P}$ in Eq.(\ref{ec_ham_qkr_2niveles})
yields
\begin{eqnarray}
\widehat{H} &=&\frac{\left[ \hbar {k_{L}}(\widehat{k}+\widehat{\beta })%
\right] ^{2}}{2M}+\hbar \Delta |e\rangle \langle e|  \notag \\
&&+K\delta _{T}(t)\cos (\widehat{\theta })(|e\rangle \langle g|+|g\rangle
\langle e|).  \label{ec_ham_qkr_2nivele}
\end{eqnarray}

It must be noted that Eq.(\ref{ec_ham_qkr_2nivele}) does not depend on the
operator $\widehat{l}$ and therefore $\widehat{\beta }$ is a preserved
quantity. Then if the initial condition belongs to a subspace corresponding
to a well defined eigenvalue of $\widehat{\beta }$, the dynamics is such
that the system remains in said subspace and the evolution of the system
will be only determined by the conjugate operators $\widehat{\theta }$ and $%
\widehat{k}$. Therefore we may restrict ourselves to the study of the
evolution constrained to a subspace corresponding to a given eigenvalue of $%
\beta$. In this case the composite Hilbert space for the Hamiltonian Eq.(\ref%
{ec_ham_qkr_2nivele}) is the tensor product $\mathcal{H} _{s}\otimes
\mathcal{H}_{c}$. $\mathcal{H}_{s}$ is the Hilbert space associated to the
discrete momentum on the line and it is spanned by the set $\{|k\rangle\}$. $%
\mathcal{H}_{c}$ is the chirality (or coin) Hilbert space spanned by two
orthogonal vectors $\{|g\rangle, |e\rangle\}$. In this composite space the
system evolves, at discrete time steps $t\in \mathbb{N}$, along a
one-dimensional lattice of sites $k\in \mathbb{Z}$. The direction of motion
depends on the state of the chirality. Taking this into account it is clear
that the Hilbert space of the 2L-QKR (with the preceding restriction) is
identical to that of the usual QW on the line.

The evolution of the system is governed by the Hamiltonian given by Eq.(\ref%
{ec_ham_qkr_2nivele}), so that, as is the case for the usual QKR, the
unitary time evolution operator for one temporal period $T$ can be written
as the application of two operators, one representing the unitary operator
due to the kick and another being the unitary operator of the free evolution
\cite{Saunders1}
\begin{equation}
\widehat{U}=e^{-i\left[ \hbar \Delta |e\rangle \langle e|+\tau (\widehat{k}+%
\widehat{\beta })^{2}\right] }e^{i\kappa cos{\widehat{\theta }}\sigma _{x}}
\label{evolu0}
\end{equation}%
where $\sigma _{x}$ is the Pauli matrix in the $x$ direction,
\begin{equation}
\tau =\frac{k_{L}^{2}\hbar }{2M}T,  \label{tau}
\end{equation}%
and
\begin{equation}
\kappa =\frac{K}{\hbar }.  \label{kappa}
\end{equation}%
The unit operator Eq.(\ref{evolu0}) in the momentum representation and in
the chirality base $\left\{ |e\rangle ,|g\rangle \right\} $ has the
following shape
\begin{eqnarray}
U(\beta )_{jk} &=&f_{jk}(\beta ,\kappa ,\tau )  \label{uu} \\
&&.\left(
\begin{array}{cc}
e^{-i\widetilde{\Delta }}\delta _{k-j\text{ }2l} & e^{-i\widetilde{\Delta }%
}\delta _{k-j\text{ }2l+1} \\
\delta _{k-j\text{ }2l+1} & \delta _{k-j\text{ }2l}%
\end{array}%
\right) ,  \notag
\end{eqnarray}%
where
\begin{equation}
f_{jk}(\beta ,\kappa ,\tau )=i^{k-j}J_{k-j}\left( \kappa \right)
e^{-i(j+\beta )^{2}\tau },  \label{efe}
\end{equation}%
$\delta _{kj}$ is the Kronecker delta, $l$ is an integer number and
\begin{equation}
\widetilde{\Delta }=T\Delta =\frac{2M}{k_{L}^{2}\hbar }\tau \Delta .
\label{delta}
\end{equation}%
The wave-vector in the momentum representation can be expressed as the
spinor
\begin{eqnarray}
|\Psi (t)\rangle &\equiv &\left(
\begin{array}{c}
|\Psi ^{e}(t)\rangle \\
|\Psi ^{g}(t)\rangle%
\end{array}%
\right)  \label{psi} \\
&=&\sum_{k=-\infty }^{\infty }\int_{-\frac{1}{2}}^{\frac{1}{2}}\left(
\begin{array}{c}
a_{k+\beta ^{\prime }}(t) \\
b_{k+\beta ^{\prime }}(t)%
\end{array}%
\right) \delta (\beta -\beta ^{\prime })|k+\beta ^{\prime }\rangle d\beta
^{\prime },  \notag
\end{eqnarray}%
where $\beta $ is the value of $\beta ^{\prime }$ for the chosen subspace
and
\begin{equation}
\left(
\begin{array}{c}
a_{k+\beta }(t) \\
b_{k+\beta }(t)%
\end{array}%
\right) =\left(
\begin{array}{c}
\langle {k+\beta }|\Psi ^{e}(t)\rangle \\
\langle {k+\beta }|\Psi ^{g}(t)\rangle%
\end{array}%
\right) ,  \label{evolu}
\end{equation}%
are the upper and lower components that correspond to the left and right
chirality of the QW.

The discrete quantum map is obtained using Eqs.(\ref{uu},\ref{psi})
\begin{equation}
\left(
\begin{array}{c}
a_{k+\beta }(t+T) \\
b_{k+\beta }(t+T)%
\end{array}%
\right) =\sum_{j=-\infty }^{\infty }U(\beta )_{kj}\left(
\begin{array}{c}
a_{j+\beta }(t) \\
b_{j+\beta }(t)%
\end{array}%
\right) .  \label{mapa}
\end{equation}%
The dynamical evolution of the system up to $t=nT$ is obtained applying the
above rule Eq.(\ref{mapa}) $n$ times.

\subsection{Resonance $\protect\tau=2\protect\pi$ in the $\protect\beta=0$
subspace with $\widetilde{\Delta }=2m\protect\pi$}

In this subsection we solve analytically the evolution of the system given
by the map Eq.(\ref{mapa}). We consider here the principal resonance $\tau
=2\pi $ in the subspace $\beta =0$. Due to the quasimomentum conservation
the value of $\beta $ does not change. Therefore the accessible momentum
spectrum is discrete and from now on the theoretical development is similar
to that of the usual QKR in resonance. Additionally we choose $\widetilde{%
\Delta }=2m\pi $ with $m$ integer in order to obtain the wave function
analytically. We will show afterwards, using numerical calculation, that the
qualitative behavior will be similar for arbitrary $\widetilde{\Delta }$.
With these conditions the matrix of Eq.(\ref{uu}) only depends on $j-k$. In
order to simplify the notation we define
\begin{equation}
U_{k_{a}k_{b}}(\kappa )=f_{k_{a}k_{b}}(0,\kappa ,2\pi )\left(
\begin{array}{cc}
\delta _{k_{a}-k_{b}\text{ }2l} & \delta _{k_{a}-k_{b}\text{ }2l+1} \\
\delta _{k_{a}-k_{b}\text{ }2l+1} & \delta _{k_{a}-k_{b}\text{ }2l}%
\end{array}%
\right) .  \label{simp}
\end{equation}%
Using Eq.(\ref{mapa}) the initial condition is connected with the wave
function at the time $t=nT$ by the equation
\begin{eqnarray}
\left(
\begin{array}{c}
a_{k_{n}}(nT) \\
b_{k_{n}}(nT)%
\end{array}%
\right) &=&\sum_{k_{n-1}}\sum_{k_{n-2}}\sum_{k_{n-3}}...  \notag \\
&&...\sum_{k_{2}}\sum_{k_{1}}\sum_{k_{0}}U_{k_{n}k_{n-1}}(\kappa
)U_{k_{n-1}k_{n-2}}(\kappa )...  \notag \\
&&...U_{k_{2}k_{1}}(\kappa )U_{k_{1}k_{0}}(\kappa )\left(
\begin{array}{c}
a_{k_{0}}^{0} \\
b_{k_{0}}^{0}%
\end{array}%
\right) ,  \label{mtotal0}
\end{eqnarray}%
where $a_{k_{0}}^{0}=a_{k_{0}}(0)$ and $b_{k_{0}}^{0}=b_{k_{0}}(0)$.

Using the relation,
\begin{eqnarray}
\sum_{k_{n-1}}U_{k_{n}k_{n-1}}(\kappa _{{1}})U_{k_{n-1}k_{n-2}}(\kappa _{{2}%
}) &=&U_{k_{n}k_{n-2}}(\kappa _{{1}}+\kappa _{{2}}),  \notag \\
&&  \label{rela0}
\end{eqnarray}%
obtained in Appendix A, Eq.(\ref{mtotal0}) is reduced to
\begin{eqnarray}
\left(
\begin{array}{c}
a_{k}(nT) \\
b_{k}(nT)%
\end{array}%
\right) &=&\sum_{j}i^{j-k}J_{j-k}\left( n\kappa \right) \left\{ \delta _{j-k%
\text{ }2l}\left(
\begin{array}{c}
a_{j}^{0} \\
b_{j}^{0}%
\end{array}%
\right) \right.  \notag \\
&&+\delta _{j-k\text{ }2l+1}\left. \left(
\begin{array}{c}
b_{j}^{0} \\
a_{j}^{0}%
\end{array}%
\right) \right\} ,\   \label{sol0}
\end{eqnarray}
where $l$ is now an arbitrary integer number.

\subsection{Antiresonance $\protect\tau=2\protect\pi$ in the $\protect\beta%
=0 $ subspace with $\widetilde{\Delta }=(2m+1)\protect\pi$}

We now find the time evolution of the wave function for $\widetilde{\Delta }%
=(2m+1)\pi$. Eq.(\ref{uu}) shows that in this case the matrix $U(\beta =
0)_{jk}$ satisfies the relation
\begin{equation}
\sum_{k_{n-1}}U_{k_{n}k_{n-1}}(\kappa )U_{k_{n-1}k_{n-2}}(\kappa )=\delta
_{k_{n}\text{ }k_{n-2}}I,  \label{rela2}
\end{equation}%
where $I$ is the identity matrix. This last expression together with Eq.(\ref%
{mapa}) imply that
\begin{eqnarray}
\left(
\begin{array}{c}
a_{k}(nT) \\
b_{k}(nT)%
\end{array}%
\right) &=&\text{\ }\delta _{n\text{ }2l+1}\sum_{j}U_{kj}(\kappa )\left(
\begin{array}{c}
a_{j}^{0} \\
b_{j}^{0}%
\end{array}%
\right)  \label{periodico} \\
&&+\text{ \ }\delta _{n\text{ }2l}\text{ }\left(
\begin{array}{c}
a_{k}^{0} \\
b_{k}^{0}%
\end{array}%
\right) .  \notag
\end{eqnarray}%
Then it is clear that the 2L-QKR shows a periodic behavior when the
parameters of the system take the values here considered. This behavior has
no analog in the usual QKR since the parameter $\widetilde{\Delta }$ does
not exist in said system. Furthermore, it is interesting to point out that
this anti-resonance occurs for $\tau=2\pi$, value for which the usual QKR is
in resonance and does not present periodic behavior.

\section{Probability distribution of momentum}

The evolution of the variance, $\sigma ^{2}=m_{2}-m_{1}^{2}$, of the
probability distribution of momentum is a distinctive feature of the QKR in
resonance. It is known that it increases quadratically in time in the
quantum case, but only linearly in the classical case. In this section we
study the evolution of the variance of the 2L-QKR, once again restricting
ourselves to the $\beta =0$ subspace and taking $\tau =2\pi$, which
corresponds to the primary resonance of the usual QKR model. We will obtain
the variance from the evolution of the first and second moments, defined as $%
m_{1}(t)=\sum_{k}kP_{k}(t)$ and $m_{2}(t)=\sum_{k}k^{2}P_{k}(t)$
respectively, where $P_{k}(t) = |a_k(t)|^2 + |b_k(t)|^2$ is the probability
to find the particle with momentum $p = \hbar k_L k$ at time $t$.

\begin{figure}[th]
\begin{center}
\includegraphics[scale=0.6, angle=0]{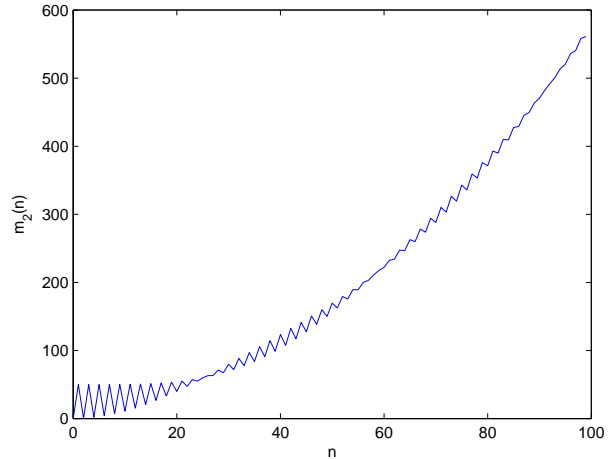}
\end{center}
\caption{The dimensionless second moment for $\widetilde{\Delta}=0.97
\protect\pi$ as a function of the dimensionless time.}
\label{f1}
\end{figure}

We first consider the resonance defined by $\widetilde{\Delta} =2m\pi $. In
this case we are able to calculate the first and second moments analytically
using Eq.(\ref{sol0}) and the properties of the Bessel functions (see
Appendix B), obtaining:
\begin{equation}
m_{1}(n)=\kappa n\sum_{j=-\infty }^{\infty }\Im \left[ a_{j}^{0}b_{j+1}^{0%
\ast }-a_{j}^{0}b_{j-1}^{0\ast }\right] +m_{1}(0)\text{,}  \label{mome0}
\end{equation}%
\begin{eqnarray}
m_{2}(n) &=&\frac{(\kappa n)^{2}}{2}\left( 1+\sum_{j=-\infty }^{\infty }\Re %
\left[ a_{j}^{0}a_{j+2}^{0\ast }(0)+b_{j}^{0}b_{j+2}^{0\ast }\right] \right)
\notag \\
&&+\kappa n\sum_{j=-\infty }^{\infty }(2j+1)\Im \left[ a_{j}^{0}b_{j+1}^{0%
\ast }+a_{j}^{0}b_{j-1}^{0\ast }\right] \text{ }  \notag \\
&&+m_{2}(0)\text{,}  \label{mome}
\end{eqnarray}%
where $\Re \left[ x\right] $ and $\Im \left[ x\right] $ are respectively the
real part and imaginary part of $x.$ $m_{1}(0)$ and $m_{2}(0)$ are the
moments at time $t=0$. These last equations show that the behavior of the
variance $\sigma^2 = m_{2}-m_{1}^{2}$ has a quadratic time dependence
irrespective of the initial conditions taken. \newline
\indent When $\widetilde{\Delta} = (2n + 1)\pi$, it was shown in the
previous section that the 2L-QKR has a periodic dynamics and therefore the
behavior of the statistical moments will be periodic as well.\newline
\indent The case when $\widetilde{\Delta} \neq 2n\pi$ is cumbersome to solve
analytically, so we restrict ourselves to a numerical study. The evolution
of the second statistical moment was obtained for different values of $%
\widetilde{\Delta}$ through numerical iterations of the map given by Eq.(\ref%
{mapa}). It was found, for all the considered values of $\widetilde{\Delta}
\neq 2n\pi$, that the long-time behavior of the second moment (and therefore
of the variance) is quadratic after an initial transient. The duration of
the initial transient depends on the initial conditions and the value of $%
\widetilde{\Delta}$. This features can be appreciated in Fig.(\ref{f1}).
The figure shows the time evolution of the second moment for the initial
conditions $|\Psi(0)\rangle = |k=0\rangle|g\rangle$. It can be appreciated
that the second moment approaches a quadratic behavior after an oscillatory
transient. It was found that the nearer the parameter $\widetilde{\Delta}$
is to $(2n+1)\pi $, the more pronounced this oscillation is.

\section{Entanglement}

\begin{figure}[th]
\begin{center}
\includegraphics[scale=0.6, angle=0]{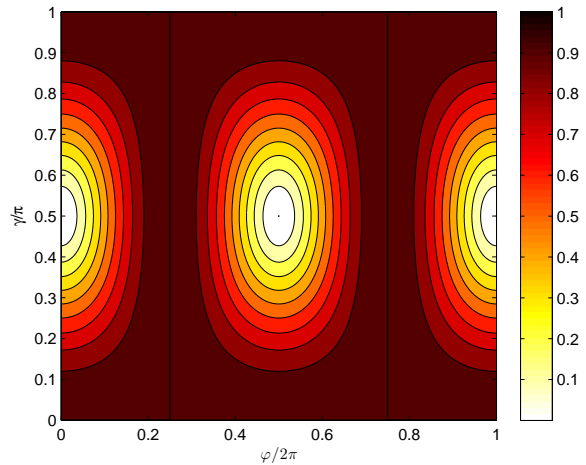}
\end{center}
\caption{The dimensionless entanglement entropy as a function of the
dimensionless initial conditions, see Eq.(\protect\ref{sol}) . The grayscale
(color online) corresponds to different values of the entropy between zero
and one.}
\label{f2}
\end{figure}

%The concept of entanglement is an important element in the development of
%quantum communication, quantum cryptography, and quantum computation.
In the context of QWs several authors \cite%
{Carneiro,Abal,Annabestani,Omar,Pathak,Liu,Venegas,Endrejat,Bracken,Ellinas,Maloyer,alejo2010,alejo2012}
have been investigating the relationship between the asymptotic
coin-position entanglement and the initial conditions of the walk. In order
to compare the model considered in this paper with the QW, we investigate
the asymptotic chirality-momentum entanglement in the 2L-QKR. The unitary
evolution of the 2L-QKR generates entanglement between chirality and
momentum degrees of freedom. This entanglement will be characterized \cite%
{Carneiro,alejo2010} by the von Neumann entropy of the reduced density
operator, called entropy of entanglement. The quantum analog of the Gibbs
entropy is the von Neumann entropy
\begin{equation}
S_{N}(\rho )=-\text{tr}(\rho \log \rho ),  \label{entropy}
\end{equation}%
where $\rho =|\Psi (t)\rangle \left\langle \Psi (t)\right\vert $ is the
density matrix of the quantum system. Owing to the unitary dynamics of the
2L-QKR, the system remains in a pure state, and this entropy vanishes. In
spite of this chirality and momentum are entangled, and the entanglement can
be quantified by the associated von Neumann entropy for the reduced density
operator:
\begin{equation}
S=-\text{tr}(\rho _{c}\log_{2} \rho _{c}),  \label{entroredu}
\end{equation}%
where $\rho _{c}=$tr$_k(\rho)$ is the reduced density matrix that results
from taking the partial trace over the momentum space. The reduced density
operator can be explicitly obtained using the wave function Eq.(\ref{psi})
in the subspace $\beta =0$ and its normalization properties
\begin{equation}
\rho _{c}=\left(
\begin{array}{cc}
P_{g}(n) & Q(n) \\
Q^{\ast }(n) & P_{e}(n)%
\end{array}%
\right) ,  \label{rc}
\end{equation}%
where
\begin{equation}
P_{g}(n)=\sum_{j=-\infty }^{\infty }\left\vert a_{k}(nT)\right\vert ^{2}%
\text{,}  \label{pe}
\end{equation}
\begin{equation}
P_{e}(n)=\sum_{j=-\infty }^{\infty }\left\vert b_{k}(nT)\right\vert ^{2}%
\text{,}  \label{peb}
\end{equation}%
\begin{equation}
Q(n)=\sum_{j=-\infty }^{\infty }a_{k}(nT)b_{k}^{\ast }(nT)\text{.}
\label{qu}
\end{equation}
$P_{e}(n)$ and $P_{g}(n)$ may be interpreted as the time-dependent
probabilities for the system to be in the excited and the ground states
respectively. In order to investigate the entanglement dependence on the
initial conditions, we consider the localized case, that is the initial
state of the rotor is assumed to be sharply localized with vanishing
momentum and arbitrary chirality, thus
\begin{equation}
\left(
\begin{array}{c}
{a_{k}(0)} \\
{b_{k}(0)}%
\end{array}%
\right) =\left(
\begin{array}{c}
\cos {\frac{\gamma }{2}} \\
\exp i\varphi \text{ }\sin {\frac{\gamma }{2}}%
\end{array}%
\right) \delta _{k0},  \label{sol}
\end{equation}%
where $\gamma \in \left[ 0,\pi \right] $ and $\varphi \in \left[ 0,2\pi %
\right] $ define a point on the unit three-dimensional Bloch sphere. Eq.(\ref%
{sol0}) takes the following form
\begin{eqnarray}
\left(
\begin{array}{c}
a_{k}(nT) \\
b_{k}(nT)%
\end{array}%
\right) &=&i^{k}J_{k}\left( n\kappa \right) \left\{ \delta _{k\text{ }%
2l}\left(
\begin{array}{c}
\cos {\frac{\gamma }{2}} \\
\exp i\varphi \text{ }\sin {\frac{\gamma }{2}}%
\end{array}%
\right) \right.  \notag \\
&&+\delta _{k\text{ }2l+1}\left. \left(
\begin{array}{c}
\exp i\varphi \text{ }\sin {\frac{\gamma }{2}} \\
\cos {\frac{\gamma }{2}}%
\end{array}%
\right) \right\} .  \label{wave}
\end{eqnarray}
Substituting Eq.(\ref{wave}) into Eqs.(\ref{pe},\ref{peb},\ref{qu}) and
using the properties of the Bessel functions, we obtain:
\begin{equation}
P_{g}(n)=\frac{1}{2}\left[1+ J_0(2n\kappa)\cos\gamma\right] \text{,}
\label{pe2}
\end{equation}%
\begin{equation}
P_{e}(n)=\frac{1}{2}\left[1- J_0(2n\kappa)\cos\gamma\right] \text{,}
\label{peb2}
\end{equation}%
\begin{equation}
Q(n)=\frac{\sin\gamma}{2}\left[\cos\varphi-i\sin\varphi J_0(2n\kappa)\right]
\text{.}  \label{qu2}
\end{equation}
The eigenvalues of the density operator $\rho_{c}$, Eq.(\ref{rc}), as a
function of $P_{g}(n)$, $P_{e}(n)$ and $Q(n)$ is
\begin{equation}
\lambda _{\pm}=\frac{1}{2}\left[ 1\pm \sqrt{1-4\left(
P_g(n)\,P_e(n)-\left\vert Q(n)\right\vert ^{2}\right) }\right],  \label{lam}
\end{equation}
and the reduced entropy as a function of these eigenvalues is
\begin{equation}
S(n)=-\lambda_{+}\log_{2} \lambda_{+}-\lambda_{-}\log_{2} \lambda_{-}.
\label{ttres}
\end{equation}
Therefore the dependence of the entropy on the initial conditions is
expressed through the angular parameters $\varphi$ and $\gamma$. This means
that, given certain initial conditions, the degree of entanglement of the
chirality and momentum degrees of freedom is determined.\newline
\indent It is seen from Eqs.(\ref{pe2},\ref{peb2},\ref{qu2}) that the
occupation probabilities and the coherence $Q$ tend to a certain limit when $%
n\rightarrow\infty$. In this limit $J_0(2n\kappa) \rightarrow0$ and both of
the occupation probabilities tend to $1/2$, irrespective of the initial
conditions. However, in the asymptotic regime, dependence on the initial
conditions is still maintained by $Q$, and therefore by the entropy as well.
Thus, in the asymptotic regime we have
\begin{equation}
\lambda _{\pm}\rightarrow\Lambda _{\pm}=\frac{1}{2}\left[ 1\pm \cos\varphi
\sin\gamma\right],  \label{lamasi}
\end{equation}
and the asymptotic value of the entropy, $S(n)\rightarrow S_{0}$, is
\begin{equation}
S_{0}=-\Lambda_{+}\log_{2} \Lambda_{+}-\Lambda_{-}\log_{2} \Lambda_{-}.
\label{ttresi}
\end{equation}
For the initial condition $\varphi= \pi/2$ and/or $\gamma= \pi$ on the Bloch
sphere, $Q\rightarrow0$ and both eigenvalues are $\Lambda_{\pm}= 1/2$. In
this case the asymptotic entanglement entropy Eq.(\ref{ttresi}) has its
maximum value $S_{0} = 1$. Finally, for sharply localized initial conditions
with zero momentum, Fig.\ref{f2} shows the dependence of the asymptotic
entanglement entropy on the parameters $\varphi$ and $\gamma$.

\section{Conclusion}

%More than ten years ago, using samples of cold atoms interacting with a
%far-detuned standing wave of laser light \cite%
%{Moore1,Robinson0,Robinson1,Ammann,Kanem}, quantum resonances and dynamical
%localization were experimentally observed. These experiments may have
%established the experimental basis of quantum computers \cite%
%{Santos,Benenti,Terraneo,Keating}. Additionally, several experiments have
%been proposed \cite{Dur,Travaglione,Sanders,Knight} to construct models of
%QWs. These proposals have some common elements with the experimental
%implementations of the QKR in the recent past.

We developed a new QKR model with an additional degree of freedom, the 2L-QKR. This system exhibits quantum resonances
with a ballistic spreading of the variance of the momentum distribution, and entanglement between the internal and
momentum degrees of freedom only depending on the initial conditions. These results were established analytically and
numerically for different values of the parameter space of this system that correspond to the primary resonance of the
usual QKR model. The above two behaviors also characterize the QW on the line and hence establish again an equivalence
between the QW and the 2L-QKR. This suggests that experiments that are related to each of two models should also carry
some kind of physical equivalence between them. We have found also that, although our system exhibits characteristics
similar to those found in the usual QKR model, there are still novel features, such as the existence of the
anti-resonance described in section II B, which have no analogue in the simple QKR model. These characteristics of the
2L-QKR render the system as an interesting candidate for further study within the framework of quantum computation.

We acknowledge stimulating discussions with V\'{\i}ctor Micenmacher, the
support from PEDECIBA and ANII.

\appendix
\section{}
Starting from Eq.(\ref{simp}) the following expression is obtained
\begin{equation}
\sum_{k_{1}}U_{k_{1}k_{2}}(\kappa )U_{k_{0}k_{1}}(\kappa
)=i^{k_{2}-k_{0}}\sum_{k_{1}}J_{\nu _{2}}\left( \kappa \right) J_{\nu
_{1}}\left( \kappa \right) \left(
\begin{array}{cc}
E_{1} & E_{2} \\
E_{3} & E_{4}%
\end{array}%
\right) \label{s0},
\end{equation}%
where%
\begin{eqnarray*}
E_{1} &=&e^{-i2\widetilde{\Delta }}\delta _{\nu _{1}\text{ }2l}\text{ }%
\delta _{\nu _{2}\text{ }2l^{\prime }}+e^{-i\widetilde{\Delta }}\left(
1-\delta _{\nu _{1}\text{ }2l}\right) \left( 1-\delta _{\nu _{2}\text{ }%
2l^{\prime }}\right) , \\
E_{2} &=&e^{-i2\widetilde{\Delta }}\delta _{\nu _{1}\text{ }2l}\text{ }%
\left( 1-\delta _{\nu _{2}\text{ }2l^{\prime }}\right) +e^{-i\widetilde{%
\Delta }}\text{ }\delta _{\nu _{2}\text{ }2l^{\prime }}\left( 1-\delta _{\nu
_{1}\text{ }2l}\right) , \\
E_{3} &=&e^{-i\widetilde{\Delta }}\text{ }\delta _{\nu _{2}\text{ }%
2l^{\prime }}\text{ }\left( 1-\delta _{\nu _{1}\text{ }2l}\right) +\delta
_{\nu _{1}\text{ }2l}\left( 1-\delta _{\nu _{2}\text{ }2l^{\prime }}\right) ,
\\
E_{4} &=&e^{-i\widetilde{\Delta }}\text{ }\left( 1-\delta _{\nu _{1}\text{ }%
2l}\right) \text{ }\left( 1-\delta _{\nu _{2}\text{ }2l^{\prime }}\right)
+\delta _{\nu _{1}\text{ }2l}\delta _{\nu _{2}\text{ }2l^{\prime }},
\end{eqnarray*}%
and with $\nu _{1}=k_{1}-k_{0}$, $\nu _{2}=k_{2}-k_{1}$. In the above equations, three different type of sums are
involved, which can be carried out using the properties of the Bessel functions (Ref.\cite{Gradshteyn}, p. 992, Eq.
\textbf{8.530}).
\begin{eqnarray}
\sum_{k_{1}}J_{k_{2}-k_{1}}\left( \kappa \right) J_{k_{1}-k_{0}}\left(
\kappa \right)  &=&J_{\mu _{2}}\left( 2\kappa \right) ,  \notag \\
&&  \label{a1}
\end{eqnarray}%
\begin{equation}
\sum_{k_{1}}J_{k_{2}-k_{1}}\left( \kappa \right) J_{k_{1}-k_{0}}\left(
\kappa \right) \delta _{k_{1}-k_{0}\text{ }2l}=  \notag
\end{equation}%
\begin{equation}
\frac{1}{2}\left[ J_{\mu _{2}}\left( 2\kappa \right) \right. \left. +\delta
_{k_{2}k_{0}}\right] ,  \label{a2}
\end{equation}%
\begin{equation*}
\sum_{k_{1}}J_{k_{2}-k_{1}}\left( \kappa \right) J_{k_{1}-k_{0}}\left(
\kappa \right) \delta _{k_{1}-k_{0}\text{ }2l}\delta _{k_{2}-k_{1}\text{ }%
2l^{\prime }}=
\end{equation*}%
\begin{equation}
\frac{1}{2}\delta _{\mu _{2}\text{ }2\left( l+l^{\prime }\right) }\left[
J_{\mu _{2}}\left( 2\kappa \right) \right. \left. +\delta _{k_{2}k_{0}}%
\right] ,  \label{a3}
\end{equation}%
where $\mu _{2}=k_{2}-k_{0}$. Substituting the above equations into Eq.(\ref{s0}) and defining $p=l+l^{\prime }$
\begin{equation}
\sum_{k_{1}}U_{k_{1}k_{2}}(\kappa _{{1}})U_{k_{0}k_{1}}(\kappa _{{2}})=\frac{%
e^{-i\widetilde{\Delta }}}{2}\left[ \left(
\begin{array}{cc}
F_{1} & F_{2} \\
F_{3} & F_{4}%
\end{array}%
\right) +\left(
\begin{array}{cc}
G_{1} & 0 \\
0 & G_{2}%
\end{array}%
\right) \right]
\end{equation}%
where%
\begin{eqnarray*}
F_{1} &=&i^{\mu _{2}}J_{\mu _{2}}\left( 2\kappa \right) \text{ }\delta _{\mu
_{2}\text{ }2p}\left( 1+e^{-i\widetilde{\Delta }}\right) , \\
F_{2} &=&i^{\mu _{2}}J_{\mu _{2}}\left( 2\kappa \right) \left( 1+e^{-i%
\widetilde{\Delta }}\right) \left( 1-\text{ }\delta _{\mu _{2}\text{ }%
2p}\right) , \\
F_{3} &=&i^{\mu _{2}}J_{\mu _{2}}\left( 2\kappa \right) \left( 1+e^{i%
\widetilde{\Delta }}\right) \left( 1-\text{ }\delta _{\mu _{2}\text{ }%
2p}\right) , \\
F_{4} &=&i^{\mu _{2}}J_{\mu _{2}}\left( 2\kappa \right) \delta _{\mu _{2}%
\text{ }2p}\left( 1+e^{i\widetilde{\Delta }}\right) , \\
G_{1} &=&\delta _{k_{2}k_{0}}\left( e^{-i\widetilde{\Delta }}-1\right) , \\
G_{2} &=&\delta _{k_{2}k_{0}}\left( e^{i\widetilde{\Delta }}-1\right) .
\end{eqnarray*}
\section{}
The probability $P_{k}(n)$ of finding the system with momentum $k$ at a time $t=nT$ is obtained using Eq.(\ref{sol0}).
\begin{equation}
P_{k}(n)=|a_{k}(n)|^{2}+|b_{k}(n)|^{2}=  \notag
\end{equation}%
\begin{equation}
=\frac{1}{2}\sum\limits_{j,l}f_{jl}[a_{j}^{0}a_{l}^{0\ast }+b_{j}^{0}b_{l}^{0\ast }]+\frac{1}{2}\sum\limits_{j,l}\Re
\left\{ f_{jl}[a_{j}^{0}b_{l}^{0\ast }]\right\}   \label{eqbess}
\end{equation}%
where%
\begin{equation}
f_{jl}=i^{l-j}\left[ J_{k-j}(n\kappa )J_{k-l}(n\kappa )+J_{k-j}(-n\kappa
)J_{k-l}(-n\kappa )\right]   \notag
\end{equation}%
and $a_{k}^{0}$ and $b_{k}^{0}$ are given by the initial conditions of the
system. To calculate the moments $m_{1}(n)$ and $m_{2}(n)$ we need the
following sums
\begin{eqnarray}
I_{jl}^{(1)} &=&i^{l-j}\sum_{k=-\infty }^{\infty }kJ_{k-j}(\kappa
)J_{k-l}(\kappa )  \notag \\
&=&j\delta _{jl}-\frac{i\kappa }{2}(\delta _{lj+1}-\delta _{lj-1}) \label{g0a}
\end{eqnarray}%
\noindent and
\begin{eqnarray}
I_{jl}^{(2)} &=&i^{l-j}\sum_{k=-\infty }^{\infty }k^{2}J_{k-j}(\kappa
)J_{k-l}(\kappa )  \notag \\
&=&\frac{\kappa ^{2}}{2}(\delta _{l}j-\frac{1}{2}(\delta _{lj+2}+\delta
_{lj-2}))  \notag \\
&&+i\kappa \lbrack \frac{1}{2}(\delta _{lj+1}+\delta _{lj-1})+j(\delta
_{lj+1}-\delta _{lj-1})]+l^{2}\delta _{jl}.  \notag \\
&&
\end{eqnarray}%
Using these expressions together with Eq.(\ref{eqbess}) and the definition of the moments we obtain the first and second
moments Eqs.(\ref{mome0},\ref{mome}).
%\begin{equation}
%m_{1}(n)=\kappa n\sum_{j=-\infty }^{\infty }\Im \left[ a_{j}^{0}b_{j+1}^{0%
%\ast }-a_{j}^{0}b_{j-1}^{0\ast }\right] +m_{1}(0)\text{,}  \label{g1}
%\end{equation}%
%and
%\begin{eqnarray}
%m_{2}(n) &=&\frac{(\kappa n)^{2}}{2}\left( 1+\sum_{j=-\infty }^{\infty }\Re %
%\left[ a_{j}^{0}a_{j+2}^{0\ast }(0)+b_{j}^{0}b_{j+2}^{0\ast }\right] \right)
%\notag \\
%&&+\kappa n\sum_{j=-\infty }^{\infty }(2j+1)\Im \left[ a_{j}^{0}b_{j+1}^{0%
%5\ast }+a_{j}^{0}b_{j-1}^{0\ast }\right] \text{ }  \notag \\
%&&+m_{2}(0).  \label{g2}
%\end{eqnarray}

\end{document}